\documentclass[ aip, jmp, amsmath,amssymb, reprint, ]{revtex4-1}
\usepackage{graphicx}
\usepackage{dcolumn}
\usepackage{bm}
\usepackage{hyperref}
\usepackage{bbm}
\usepackage[usenames]{color}
\begin{document}


\title{Time-delayed feedback control of coherence resonance near subcritical Hopf bifurcation: theory versus experiment}

\author{Vladimir Semenov}
\affiliation{ Department of Physics, Saratov State University, Astakhanskaya str. 83, 410012 Saratov, Russia}
\author{Alexey Feoktistov}
\affiliation{ Department of Physics, Saratov State University, Astakhanskaya str. 83, 410012 Saratov, Russia}
\author{Tatyana Vadivasova}
\affiliation{ Department of Physics, Saratov State University, Astakhanskaya str. 83, 410012 Saratov, Russia}
\author{Eckehard Sch\"oll}
\email{schoell@physik.tu-berlin.de}
\affiliation{ Institut f\"ur Theoretische Physik, Technische Universit\"at Berlin, Hardenbergstr. 36, 10623 Berlin, Germany}
\author{Anna Zakharova}
\email{anna.zakharova@tu-berlin.de}
\affiliation{ Institut f\"ur Theoretische Physik, Technische Universit\"at Berlin, Hardenbergstr. 36, 10623 Berlin, Germany}

\date{\today}

\begin{abstract}
Using the model of a generalized Van der Pol oscillator in the regime of subcritical Hopf bifurcation we investigate the influence of time delay on noise-induced oscillations. It is shown that for appropriate choices of time delay either suppression or enhancement of coherence resonance can de achieved. Analytical calculations are combined with numerical simulations and experiments on an electronic circuit.
\end{abstract}

\maketitle

\begin{quotation}
The generalized Van der Pol oscillator is a paradigmatic model in nonlinear dynamics and electrical engineering. In contrast to a standard Van der Pol model, the generalized Van der Pol oscillator allows to describe both supercritical and subcritical Hopf bifurcations. In the regime close to a saddle-node bifurcation of periodic orbits (subcritical case) under the influence of noise the model demonstrates the phenomenon of coherence resonance. It means that there exists an optimum intermediate value of the noise intensity for which noise-induced oscillations become most coherent. Using experiments on an electronic circuit, numerical simulations, and an analytical approach, we show how to control the system with subcritical Hopf bifurcation by time-delayed feedback. In particular, we demonstrate experimentally that time delay allows both enhancement and suppression of coherence resonance.
\end{quotation}

\section{Introduction}
The phenomenon of coherence resonance\cite{HU93a, PIK97, LIN99a,LIN04} was originally discovered for excitable systems. It implies that noise-induced oscillations become more regular for an optimum value of noise intensity. These oscillations can be also synchronized mutually as well as by an external forcing\cite{HAN99, CIS03, CIS04}. Moreover, the synchronization of noise-induced oscillations occurs in a similar way as for a deterministic quasiperiodic system\cite{AST11}. It has been shown that coherence resonance can be modulated by applying time-delayed feedback in systems with type-I\cite{AUS09} and type-II\cite{JAN04} excitability. 
 
Coherence resonance has also been found in non-excitable systems with a subcritical Hopf bifurcation\cite{USH05,ZAK10a,FEO12,ZAK13,GEF14}. It is important to note that the pure coherence resonance effect for non-excitable systems is observed for a subcritical Hopf bifurcation and not for the supercritical case. The standard Van der Pol model close to a supercritical Hopf bifurcation has been investigated in the presence of delay and noise\cite{JAN04,SCH04b,POM05a}, but the interplay of noise and delay with respect to coherence resonance in the subcritical Van der Pol system has not been considered. In the present work we aim to study coherence resonance in the generalized Van der Pol system with a subcritical Hopf bifurcation. We demonstrate theoretically that coherence resonance can be modulated by time-delayed feedback and confirm our results in experiment with an electronic circuit. We consider the regime close to the saddle-node bifurcation of periodic orbits, where in the deterministic case the only attractor of the system is a stable focus. The oscillations are induced by noise and further controlled by time delay. The importance of this issue is emphasized by the fact that delay and noise are very often invoked not only in theoretical investigations \cite{GAU09,GAU10,GAU12,KOU10a,LAF13}, but also in real-world applications. For example, coherence resonance appears also in microwave dynamical systems such as a five-cavity delayed-feedback klystron oscillator at the self-excitation threshold\cite{DMI11}, in lasers with optical feedback \cite{GIA00,AVI04,OTT14a}, or in semiconductor superlattices \cite{HIZ08b,HUA13a}. 
In general, time delay arises inevitably in every real device due to final signal propagation velocity. Delayed feedback is a powerful tool for achieving a wide range of operating regimes and enhancing amplitude-frequency characteristics, and
controlling the stochastic or deterministic dynamics of nonlinear systems \cite{SCH07,SUN13,FLU13}.

\section{model}
\begin{figure*}
\begin{center}
\includegraphics[width=1.5\columnwidth]{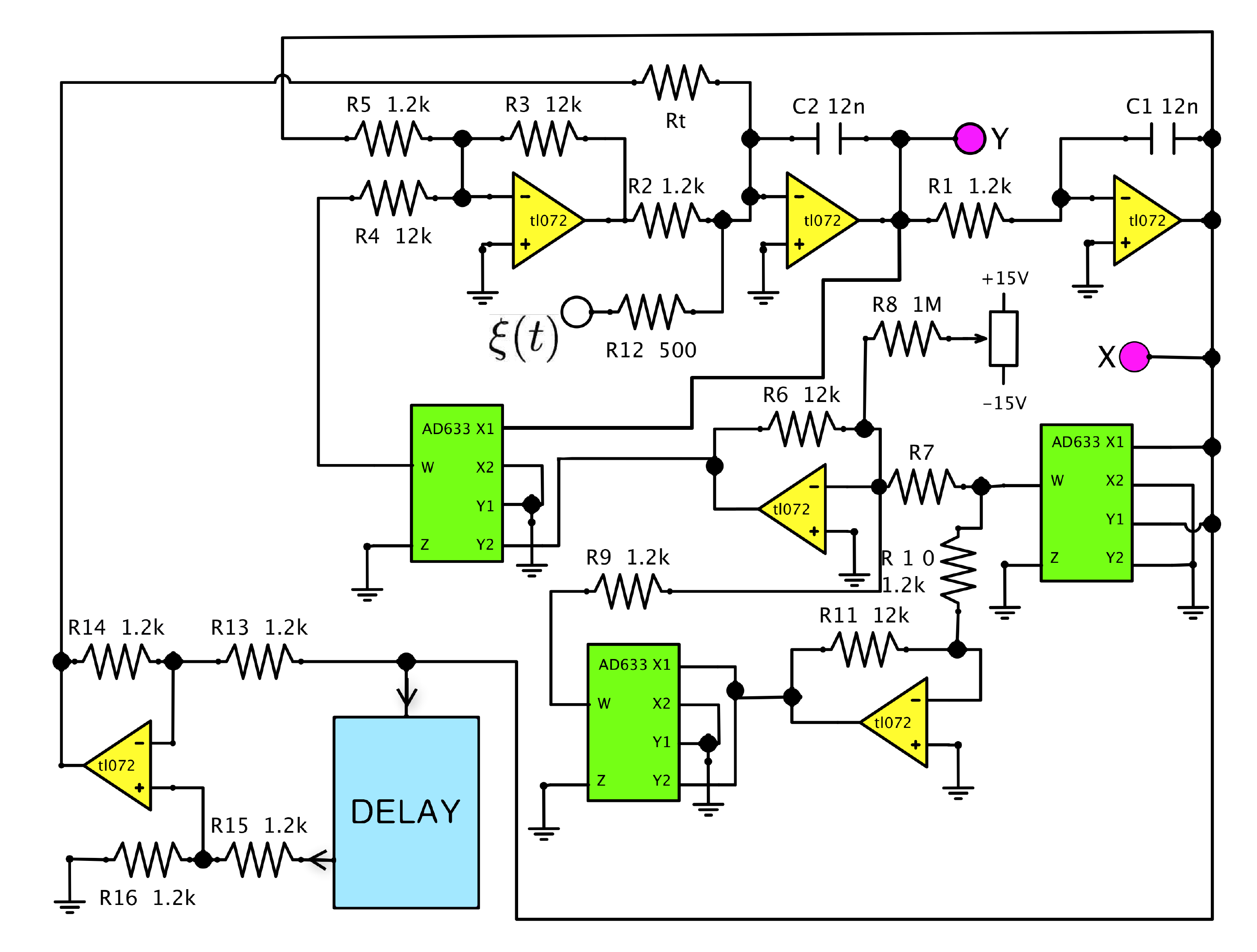}
\end{center}
\caption{Scheme of experimental setup: electronic circuit}
\label{fig1}                                                                                                   
\end{figure*}

We investigate the generalized Van der Pol oscillator, extended by a quartic term in the nonlinear friction.
If we additionally involve noise and time-delayed feedback, it is described by the following equation:
\begin{equation}
\begin{array}{l}
\dfrac{d^2x}{dt^2} - \bigg[\varepsilon +\mu x^2 -x^4 \bigg]\dfrac{dx}{dt} + \omega^2_0x =\\
\\
=\sqrt{2D}\xi(t) + K(x(t-\tau) - x(t)),
\end{array}
\label{eq1}
\end{equation}
where $x$ is the dimensionless variable, $t$ is the dimensionless time, $\varepsilon \in \mathbbm{R}$ and $\mu >0$ are the parameters responsable for excitation and dissipation, respectively, $\omega_0$ is the eigenfrequency of linear oscillations at the Hopf bifurcation, $K$ is the strength of time-delayed feedback, $\tau$ is the delay time, $\xi(t)$ is normalized Gaussian white noise: $\langle \xi(t)\xi(t+\tau) \rangle = \delta(\tau)$, $\langle \xi(t) \rangle = 0$, and $D$ is the noise intensity. 
Because of the quartic nonlinearity the system can exhibit simultaneously two limit cycles: a stable and an unstable one. The regime of coexistence of these two periodic orbits is limited from one side by a saddle-node bifurcation of limit cycles ($\varepsilon = -\mu^2/8$), and from the other side by a subcritical Hopf bifurcation ($\varepsilon = 0$). Besides two limit cycles, the regime $ -\mu^2/8 < \varepsilon < 0$ contains a stable focus in the origin. For $ \varepsilon < -\mu^2/8 $ the only attractor is the stable focus. It becomes unstable for $ \varepsilon > 0 $ in a subcritical Hopf bifurcation at $ \varepsilon = 0 $. 

For the experimental confirmation of our theoretical results we consider an electronic circuit, which models the system Eq. (\ref{eq1}). The corresponding scheme of the experimental setup is shown in Fig. \ref{fig1}. 
The setup consists of two main parts: an analog and a digital one. The analog part models the generalized Van der Pol oscillator, while the digital one provides time delay. The analog part contains operational amplifiers, analog multipliers, and additional  passive elements. The digital part consists of microcontroller ATmega16 and external analog-to-digital and digital-to-analog converters. The scheme in Fig. \ref{fig1} is described by the following equation:
\begin{equation}
\begin{array}{l}
\dfrac{d^2v_1}{dt^2} - \dfrac{1}{RC}\bigg[v_{\lambda} +\dfrac{R}{R_7} v^2_1 -v^4_1 \bigg]\dfrac{dv_1}{dt} + \dfrac{1}{(RC)^2}v_1 = \\
\\
=\dfrac{1}{(RC)^2}v_{\xi} +\dfrac{R}{R_{\tau}(RC)^2}(v_{1\tau} - v_1).
\end{array}
\label{eq2-2}
\end{equation}
Here $v_1$ is the voltage at the point $x$ on the scheme (Fig. \ref{fig1}), $v_{1\tau}=v_1(t-\tau)$, and $v_{\xi}$ is the noise voltage.
The values of the resistors and capacitors are chosen such that $R_1=R_2=R_4=R_9=R_{10}=R_{12}=R_{13}=R_{14}=R_{15}=R_{16}\equiv R$, $R_3=R_5=R_6=R_8=R_{11}=10R$ and $C_1=C_2 \equiv C$. 
In dimensionless form and with the substitution $x = v_1$, $t'=\omega t$, $\tau'=\omega \tau$, $\omega=1/(RC)$, $\varepsilon = v_{\lambda}$, $\mu=R/R_7$, $K=R/R_{\tau}$ and $\sqrt{2D}\xi=v_{\xi}$ 
Eq. (\ref{eq2-2}) is reduced to Eq. (\ref{eq1}) with $\omega_0=1$, where we have dropped the $'$ in the dimensionless time. Further details  on this  experimental setup without time delay, including a bifurcation diagram, can be found in Ref. \cite{ZAK13}. It is important to note that despite of the direct correspondence between Eq. (\ref{eq2-2}), which describes the experimental setup, and Eq. (\ref{eq1}), the experimental values of the bifurcation points differ from the values obtained numerically and analytically. This is due to the fact that Eq. (\ref{eq2-2}) was derived using standard approximations on operation amplifiers, which are common in electronics. Since our main goal is to confirm the observed phenomena qualitatively and not quantitatively, this difference is not essential for our study.

\section{Noise-free system}
We begin our investigations with the case of a noise-free system, since it is important  to understand how the time-delayed feedback influences the deterministic system. The model equation reduces to:
\begin{equation}
\begin{array}{l}
\dfrac{d^2x}{dt^2} - \bigg[\varepsilon +\mu x^2 -x^4 \bigg]\dfrac{dx}{dt} + \omega^2_0x 
=K(x(t-\tau) - x(t)).
\end{array}
\label{eq2}
\end{equation}

We assume that the amplitude of the oscillations is changing slowly on the time-scale of the period of oscillation. Then we can apply the averaging method (quasiharmonic reduction) to the Van der Pol equation. To find the solution of Eq.(\ref{eq2}) we use the following ansatz:
\begin{equation}
x(t)=Re\{A(t)\exp(i \omega_0 t)\}=\dfrac{1}{2}\{A\exp(i \omega_0 t)+ c.c.\},
\label{eq3}
\end{equation}
where $A(t)$ is a complex amplitude, and $c.c.$ denotes the complex conjugate $A^*\exp(-i \omega_0 t)$.
The first and second derivatives 
\begin{equation}
\dfrac{dx}{dt}=\dfrac{1}{2}\left[\left(\dfrac{dA}{dt}+i\omega_0 A\right)\exp(i\omega_0t) + c.c.\right],
\label{eq4}
\end{equation}
\begin{equation}
\dfrac{d^2x}{dt^2}=\dfrac{1}{2}\left[\left(\dfrac{d^2A}{dt^2} + 2i\omega_0\dfrac{dA}{dt}-\omega_0^2 A\right)\exp(i\omega_0t)+c.c.\right].
\end{equation}
can be approximated by
\begin{equation}
\dfrac{dx}{dt}=\dfrac{1}{2}\left[i\omega_0 A \exp(i\omega_0t) + c.c.\right],
\label{eq5}
\end{equation}
and 
\begin{equation}
\dfrac{d^2x}{dt^2}=\dfrac{1}{2}\left[\left(2i\omega_0\dfrac{dA}{dt} - \omega_0^2 A\right)\exp(i\omega_0t)+c.c.\right].
\label{eq6}
\end{equation}
for slowly varying amplitude $\left|\dfrac{dA}{dt}\right|\ll \omega_0 A$.

\begin{figure}
\begin{center}
\includegraphics[scale=0.35]{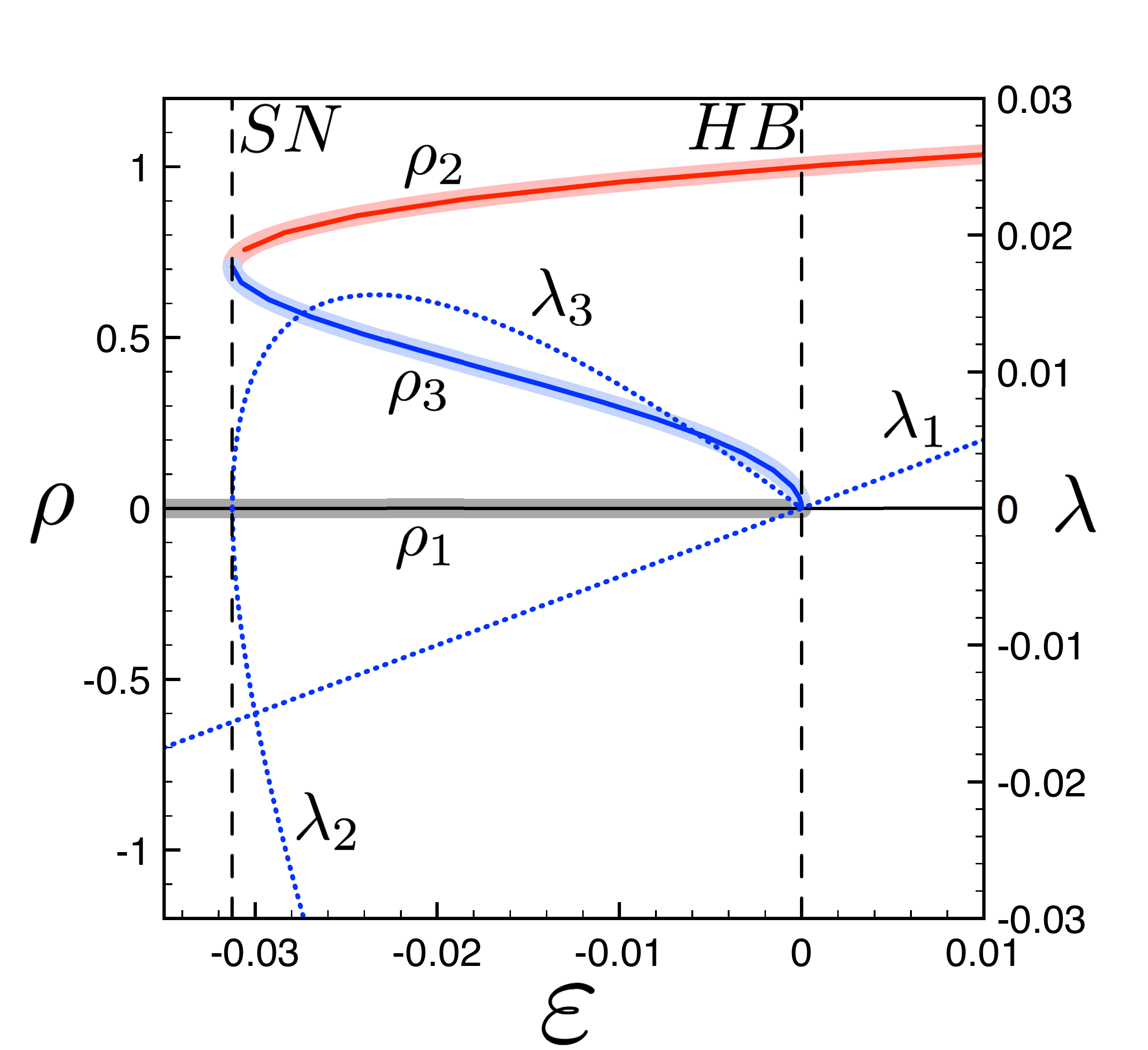}
\end{center}
\caption{Bifurcation diagram of the generalized Van der Pol oscillator without noise and delay. Analytical solutions of Eq. (\ref{eq11}): red shadowing corresponds to the stable limit cycle $\rho_{2}$, blue shadowing shows the unstable limit cycle $\rho_{3}$, black shadowing represents the steady state $\rho_1$. Solid lines stand for the simulations performed with XPPAUT. The eigenvalues are shown by blue dotted lines. The vertical dashed lines mark the saddle-node (SN) and the Hopf bifurcation (HB). Parameter values are: $K=0, D=0, \mu=0.5, \omega_0=1$. }
\label{fig2}                                                                                                   
\end{figure}

Substituting Eq. (\ref{eq3}), Eq. (\ref{eq5}), Eq. (\ref{eq6}) into Eq. (\ref{eq2}), we approximate all fast oscillating terms
($\exp(3i\omega_0t), \exp(5i\omega_0t)$ and c.c.) by their averages over one period $T=2\pi/\omega_0$ which gives zero. Furthermore we assume that the delay $\tau$ is small, so that we can approximately set $A(t-\tau)\approx A(t)$ on the slow time scale of $A(t)$.
Then we obtain
\begin{equation}
\begin{array}{l}
\dfrac{dA}{dt}=\dfrac{1}{2}\varepsilon A + \dfrac{1}{8}\mu |A|^2A - \dfrac{1}{16}|A|^4 A +\\
\\
+ \dfrac{1}{2}\dfrac{K}{\omega_{0}}A(i(1-\cos(\omega_{0}\tau)) - \sin(\omega_{0}\tau)).
\end{array}
\label{eq7}
\end{equation}
In order to solve Eq. (\ref{eq7}), we transform to polar coordinates.
 \begin{equation}
A=\rho \exp(i\phi),
\label{eq8}
\end{equation}
where $\rho \ge 0$ is the amplitude and $\phi \in \mathbbm{R}$ is the phase of oscillations. After substituting Eq. (\ref{eq8}) into Eq. (\ref{eq7}) and separating the resulting equation into real and imaginary parts, we obtain:
 \begin{equation}
 \dfrac{d\rho}{dt}=[ \dfrac{1}{2}\varepsilon + \dfrac{1}{8}\mu \rho^2 - \dfrac{1}{16}\rho^4 - \dfrac{1}{2 \omega_{0}}K\sin(\omega_{0}\tau)]\rho, \\
\label{eq9}
\end{equation}
\begin{equation}
\dfrac{d\phi}{dt}=\dfrac{1}{2\omega_{0}} K(1-\cos(\omega_{0}\tau)),
\label{eq10}
\end{equation}
where Eq. (\ref{eq9}) describes the amplitude and Eq. (\ref{eq10}) the phase dynamics. Using the conditions $\dot{\rho}=0, \dot{\phi}=0$ we find the steady state. We are only interested in stationary solutions of the amplitude equation, which describe steady states or limit cycles:
 \begin{equation}
[\dfrac{1}{2}\varepsilon + \dfrac{1}{8}\mu \rho^2 - \dfrac{1}{16}\rho^4 - \dfrac{1}{2\omega_{0}}K\sin(\omega_{0}\tau)]\rho=0.
\label{eq11}
\end{equation}
Equation (\ref{eq11}) has three solutions $\rho \ge 0$.
The substitution $\alpha=2\mu$ and $\beta= 8(\varepsilon -\dfrac{K}{\omega_{0}}sin(\omega_{0}\tau))$ 
gives the solution in the following form:
 \begin{equation}
\rho_1=0,
\label{eq12}
\end{equation}
\begin{equation}
\rho_2=\sqrt{\dfrac{\alpha+\sqrt{\alpha^2+4\beta}}{2}},
\label{eq13}
\end{equation}
\begin{equation}
\rho_3=\sqrt{\dfrac{\alpha-\sqrt{\alpha^2+4\beta}}{2}},
\label{eq15}
\end{equation}
Further we perform a linear stability analysis of these solutions for $K=0$.
The solution Eq. (\ref{eq12}) is characterized by the eigenvalue: 
\begin{equation}
\lambda_1 =\dfrac{1}{2}\varepsilon.
\label{eq23}
\end{equation}
Therefore, it is stable for $\varepsilon < 0$ and unstable for $\varepsilon > 0$.
The solution $\rho_{2}$ Eq. (\ref{eq13}) exists for $-\dfrac{\mu^2}{8} \le \varepsilon < \infty$. In this regime the stability of the solution $\rho_{2}$ is given by the eigenvalues 
\begin{equation}
\lambda_{2} =-2\varepsilon - \dfrac{1}{4}\mu^2 - \dfrac{1}{4}\mu\sqrt{\mu^2+8\varepsilon},
\label{eq21}
\end{equation}
which is always negative (Fig. \ref{fig2}). It means that the solution $\rho_{2}$ corresponds to a stable limit cycle in the system Eq.(\ref{eq2}). The solution $\rho_{3}$  Eq. (\ref{eq15}) represents an unstable limit cycle in the system Eq. (\ref{eq2}) and exist for $\varepsilon \in[-\dfrac{\mu^2}{8},0]$ where it always has a positive eigenvalue
\begin{equation}
\lambda_{3} =-2\varepsilon - \dfrac{1}{4}\mu^2 + \dfrac{1}{4}\mu\sqrt{\mu^2+8\varepsilon}.
\label{eq22}
\end{equation}
All the solutions and eigenvalues as a functions of $\varepsilon$ are shown in Fig. \ref{fig2}. 

\begin{figure}
\begin{center}
\includegraphics[scale=0.35]{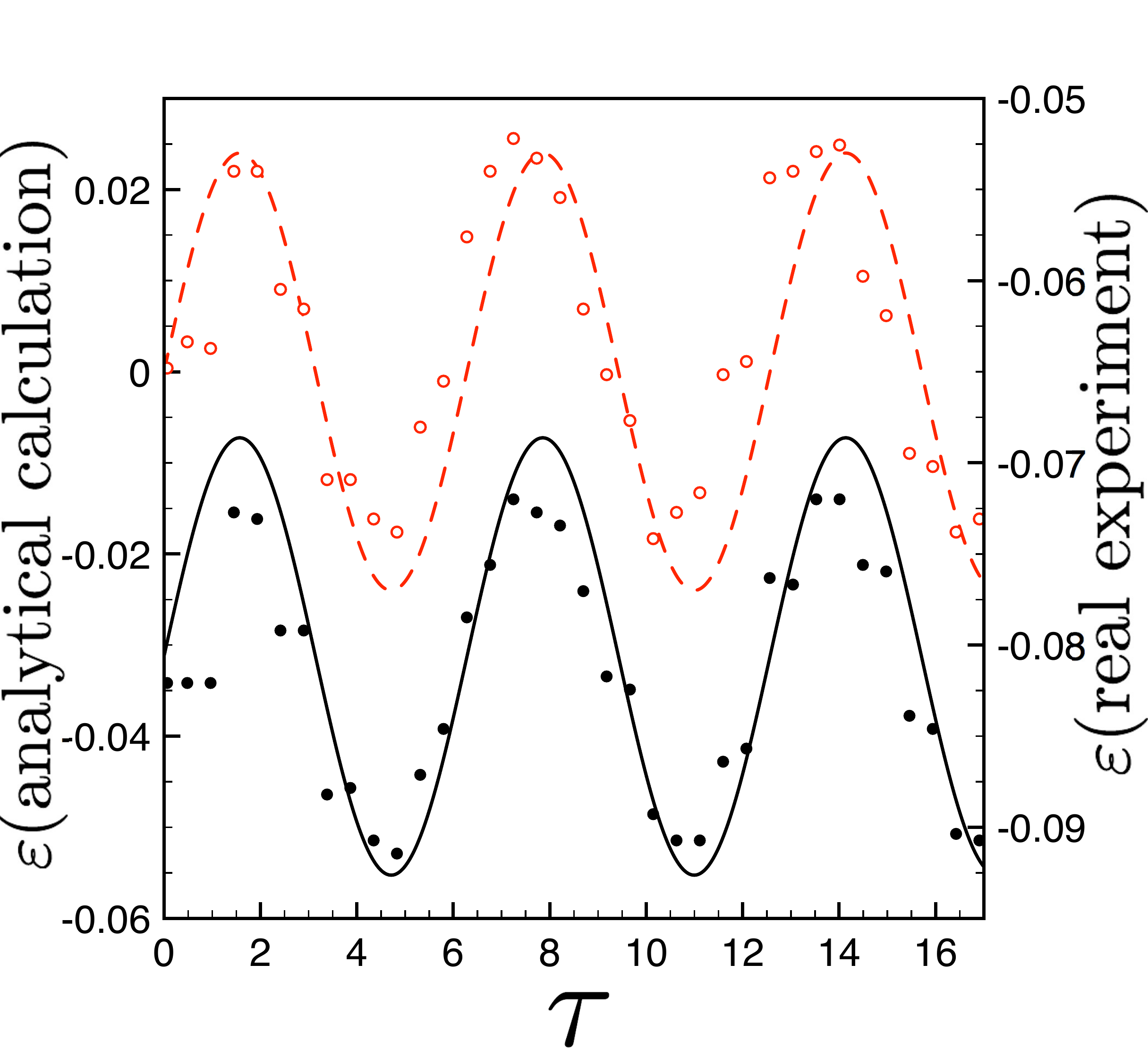}
\end{center}
\caption{Modulation of the bifurcation parameter $\varepsilon$ by time delay $\tau$: Analytical result and electronic circuit experiment. The saddle-node bifurcation of limit cycles ($\varepsilon (\tau=0)=-0.03125$) is represented by red dashed lines (analytical) and red open circles (experiment). The subcritical Hopf bifurcation ($\varepsilon(\tau=0)=0$) corresponds to black solid lines (analytical) and black dots (experiment). Parameters: $K=0.024$, $D=0$, $\mu=0.5$, $\omega_0=1$.}
\label{fig3}                                                                                                   
\end{figure}

If delay is added in the slowly-varying-amplitude approximation, this simply amounts to a rescaling of the bifurcation parameter
$\varepsilon \to \varepsilon -\dfrac{K}{\omega_{0}}sin(\omega_{0}\tau))$
The two periodic orbits collide in a saddle-node bifurcation, which implies $\rho_2=\rho_3$. From this condition one can derive $\varepsilon$ as a function of $\tau$ at the saddle-node bifurcation of limit cycles: 
\begin{equation}
\varepsilon(\tau)= -\dfrac{\mu^2}{8}+\dfrac{K}{\omega_{0}}\sin(\omega_{0}\tau).
\label{eq17}
\end{equation}
Fixing $K=0, \mu=0.5$ gives $\varepsilon=-0.03125$, in agreement with the value obtained numerically with the continuation tool XPPAUT.
The unstable limit cycle $\rho_{3}$ and the stable fixed point $\rho_1$ collide in the subcritical Hopf bifurcation at 

\begin{equation}
\varepsilon(\tau)= \dfrac{K}{\omega_{0}}\sin(\omega_{0}\tau).
\label{eq18}
\end{equation}
It follows that for $\tau \ne 0$ the maximum shift of the bifurcation parameter $\varepsilon$ from its value without delay is observed for $\omega_0\tau=\dfrac{\pi}{2}+\pi n$, $n \in \mathbbm{N}$. The dependence of $\varepsilon$ on $\tau$ obtained analytically from Eqs.~(\ref{eq17}),(\ref{eq18}) agrees well with the experimental measurements (Fig. \ref{fig3}). 

\section{Noise-induced oscillations}
\begin{figure}
\begin{center}
\includegraphics[scale=0.35]{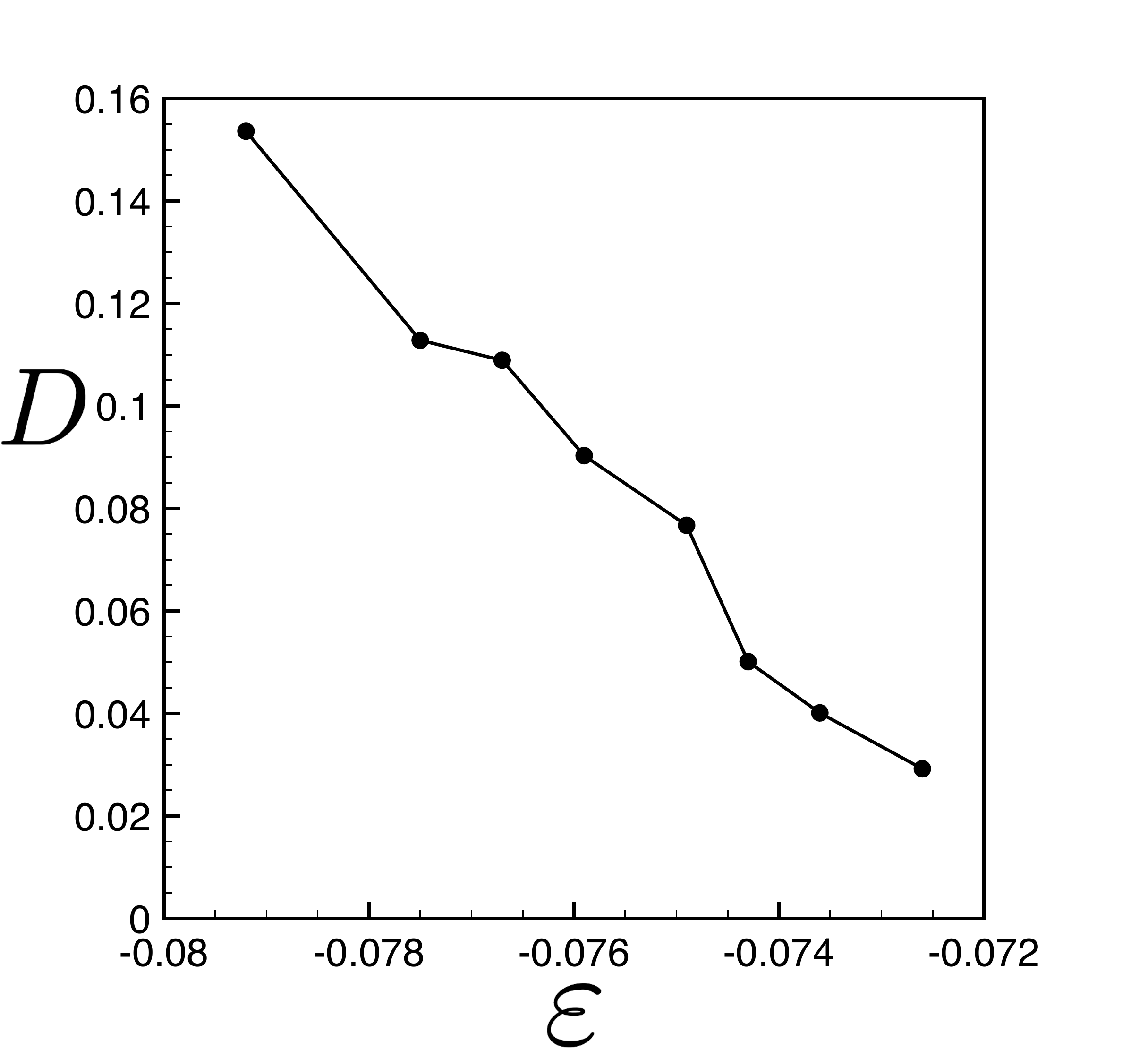}
\end{center}
\caption{Dependence of the optimum noise intensity on the bifurcation parameter $\varepsilon$ (experiment). Parameters: $K=0$, $\mu=0.5$, $\omega_0=1$.}
\label{figDL}                                                                                                   
\end{figure}

\begin{figure}
\begin{center}
\includegraphics[scale=0.35]{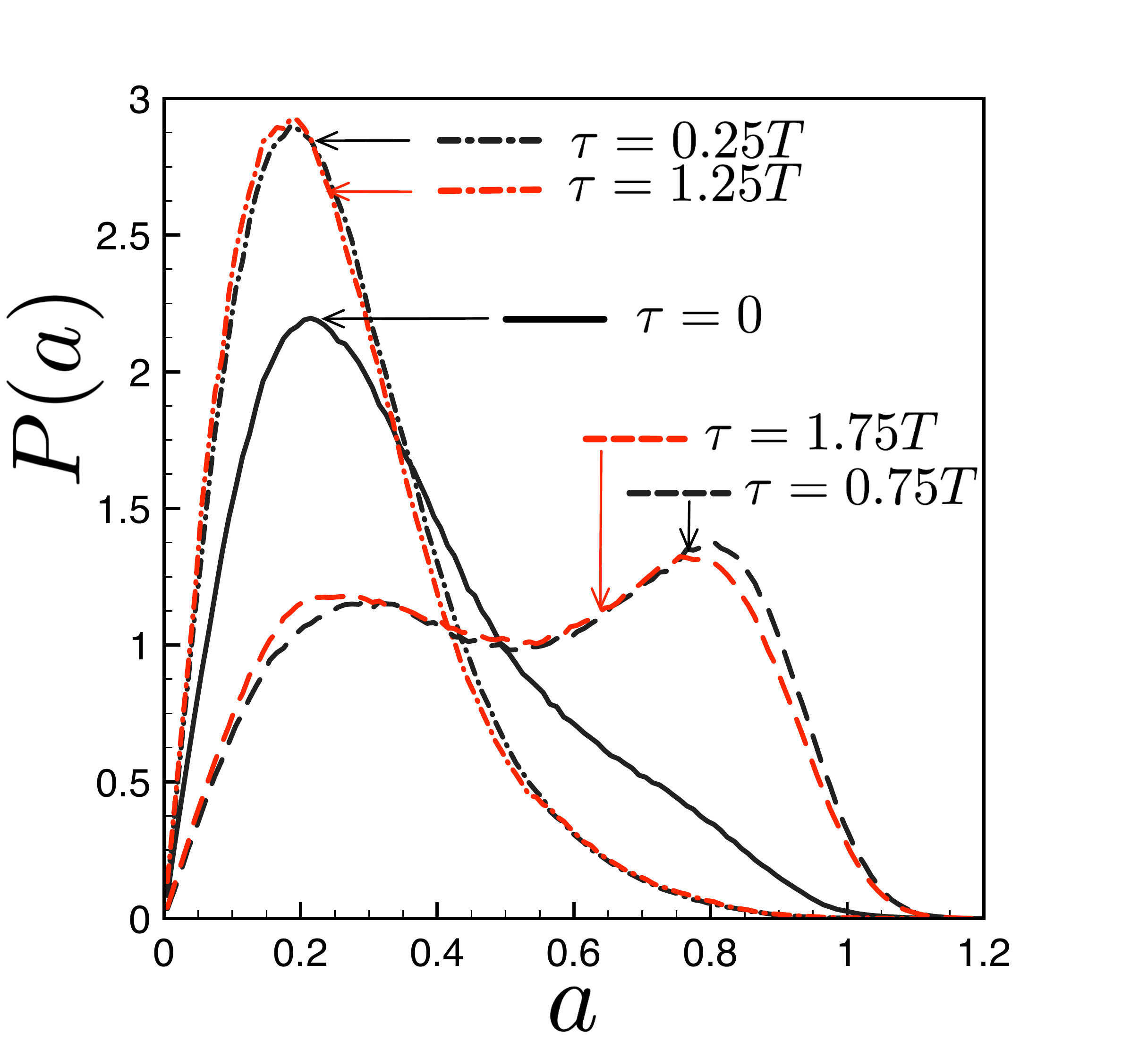}
\end{center}
\caption{Experimental results for the probability density distribution of the amplitude for various values of $\tau$. Parameters: $K=0.024$, $D=0.01$, $\mu=0.5$, $\varepsilon = -0.093$, $\omega_0=1$, $T=2\pi$.}
\label{fig8}                                                                                                   
\end{figure}

\begin{figure}
\begin{center}
\includegraphics[scale=0.35]{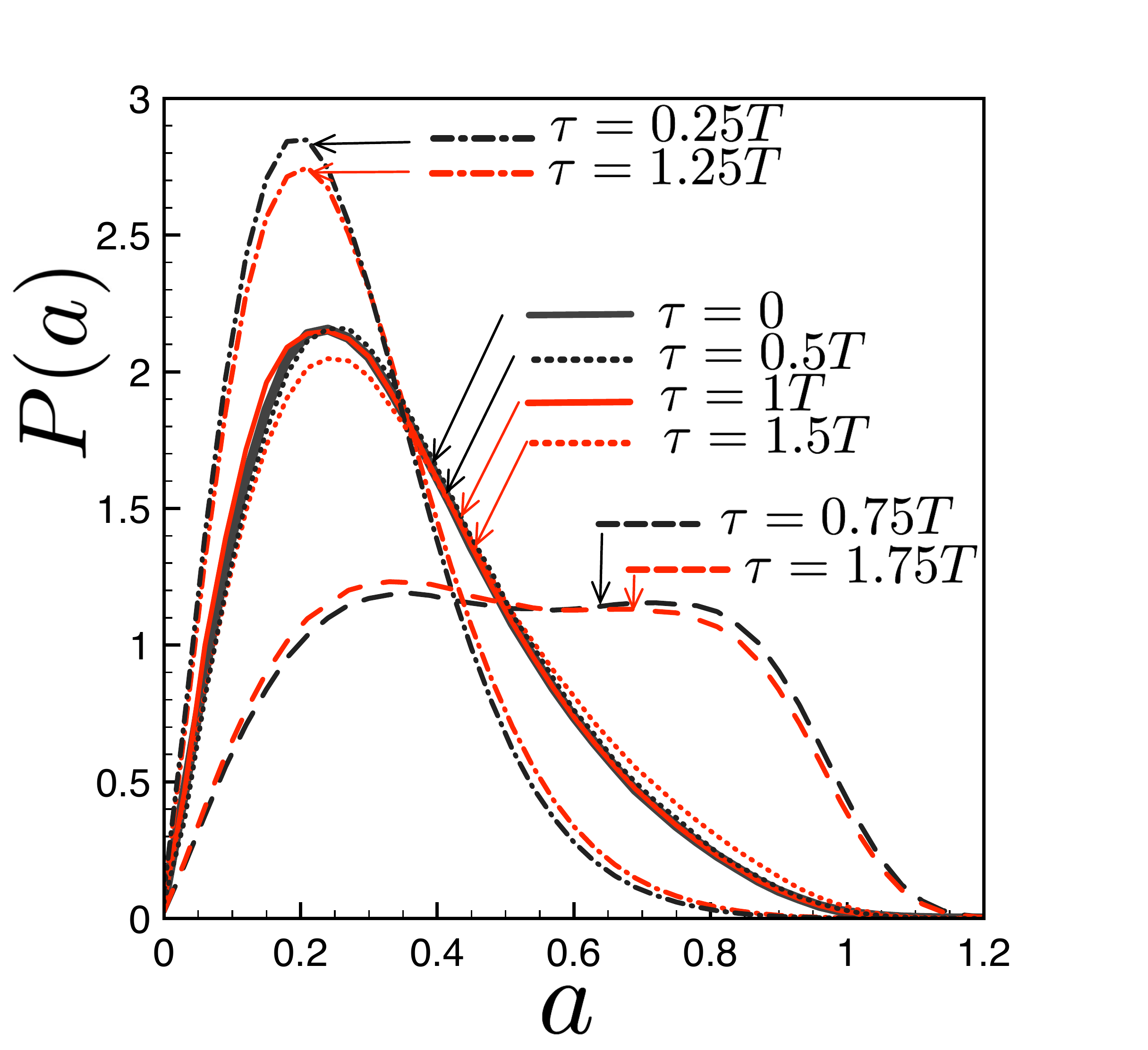} \\
\end{center}
\caption{Numerical simulation of the probability density distribution of the amplitude for various values of $\tau$. Parameters: $K=0.024$, $D=0.003$, $\mu=0.5$, $\varepsilon = -0.06$, $\omega_0=1$, $T=2\pi$.}
\label{fig9}                                                                                                   
\end{figure}

From the analysis of the system Eq. (\ref{eq2}) in the absence of noise we now know that time-delayed feedback can shift the
bifurcation point in both directions, which also can cause the change of the dynamical regime. Moreover, it is known how the distance between the operating point and the saddle-node bifurcation point influences the optimum value of the noise intensity for the coherence resonance:  the larger the distance, the stronger is the noise required to achieve coherence resonance (Fig. \ref{figDL}). Thus, by tuning the time delay value we can control the coherence resonance.
Recent works have uncovered the link between coherence resonance and stochastic $P$-bifurcation\cite{ZAK10a,ZAK13}.  This type of bifurcation appears when the variation of the noise characteristics (like noise intensity) causes a qualitative change of the probability density distribution, e.g., from monomodal to bimodal. Experimental evidence of the P-bifurcation in an electronic circuit is reported in Ref. \cite{FEO12}. Moreover, it is shown experimentally that the bimodal probability distribution corresponds to the most pronounced coherence resonance\cite{ZAK13}. Therefore, in the following we use the probability density distribution for the study of the impact of time delay upon coherence resonance. For the investigation of the system Eq. (\ref{eq1}) we choose the resonance value of the noise intensity, and choose the time-delayed feedback strength $K=0.024$. The generalized van der Pol equation (\ref{eq1}) can be re-written as a 2-variable dynamical system:
\begin{equation}
\label{eq1-tr}
\begin{array}{l}
\dfrac{dx}{dt} = y, \\
\dfrac{dy}{dt} = [\varepsilon +\mu x^2 -x^4 ] y - \omega^2_{0}x + \sqrt{2D}\xi(t) + K(x(t-\tau) - x(t)).
\end{array}
\end{equation}
The probability distribution of the amplitude is calculated from the numerical simulation of Eq. (\ref{eq1-tr}), where the amplitude is defined as $a(t)=\sqrt{x^{2}(t)+y^{2}(t)}$. In experiment, the amplitude distribution is obtained in a similar way from the measured $x(t)$ and $y(t)$.
We analyze the probability density distribution of the amplitude for varying time delay. The results of the experiment (Fig. \ref{fig8}) and the numerical simulation (Fig. \ref{fig9}) for the corresponding probability distributions agree well. 
It can be seen that for $\tau = T/4$, where $T=2 \pi/\omega_0$, the distribution is unimodal. This means that the trajectory is fluctuating in phase space near the deterministic stable focus and there is no coherence resonance. For $\tau = 3T/4$ the distribution has a well-pronounced bimodal shape, which indicates the appearance of noise-induced oscillations with a non-zero amplitude. The maximum of the distribution close the origin is due to the noisy motion near the stable focus and the second maximum corresponds to the amplitude of the noise-induced oscillatory motion. This behavior repeats with a period $T$. 
For delay times $\tau = 0, T/2, T, 3T/2$ the distribution is practically not changed, and is still unimodal. Therefore, if the delay time is equal to a multiple of half the intrinsic period of deterministic oscillations, the system behaves in the same way as without delay.  

\begin{figure}
\begin{center}
\includegraphics[scale=0.35]{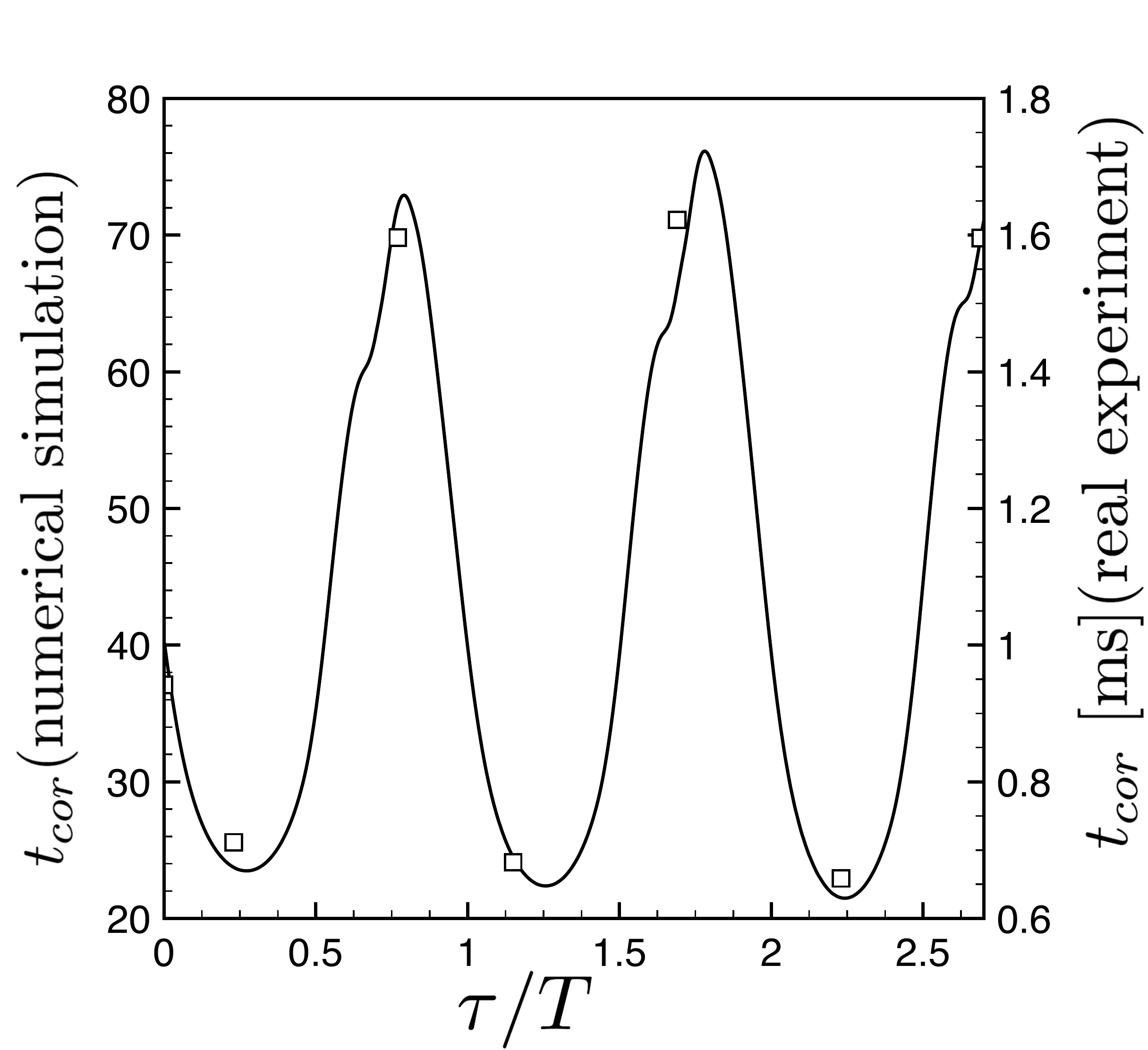}
\end{center}
\caption{Both numerically obtained (solid line) and measured (open squares) dependences of correlation time on time delay. Parameters: $K=0.024$, $\mu=0.5$, $\varepsilon = -0.06$, $\omega_0=1$, $T=2\pi$, $D=0.003$ (numerical experiment), $D=0.01$ (real experiment).}
\label{fig11}                                                                                                   
\end{figure}

To investigate the impact of time delay on coherence resonance we also calculate the correlation time $t_{cor}$ defined by
$t_{cor}=\frac{1}{\sigma^{2}}\int_{0}^{\infty} |\Psi(s)| ds$, where $\Psi(s)=\langle x(t)x(t+s)\rangle$ is the autocorrelation function, and $\sigma^{2}=\langle x(t)^2\rangle$ is the variance. It is approximately related to the full width at half maximum (FWHM) of the power spectral density $\Delta \omega$ by $t_{cor}=\frac{4}{\pi\Delta \omega}$ \cite{SCH04b,ZAK13}.
The correlation time is a common measure used for the diagnostics of coherence resonance. Again we achieve good agreement of experimental results with numerical simulations (Fig. \ref{fig11}). The dependence of the correlation time $t_{cor}$ on the time delay $\tau$ has well-pronounced minima (Fig. \ref{fig11}), which correspond to a unimodal probability density distribution (Fig. \ref{fig8},  \ref{fig9}) and appear for $\tau = (\frac{1}{4}+n)T$, $n\in \mathbb N$. The maxima are related to bimodal probability distributions of the amplitude and observed for $\tau = (\frac{3}{4}+n)T$, $n\in \mathbb N$.

\begin{figure}
\begin{center}
\includegraphics[scale=0.35]{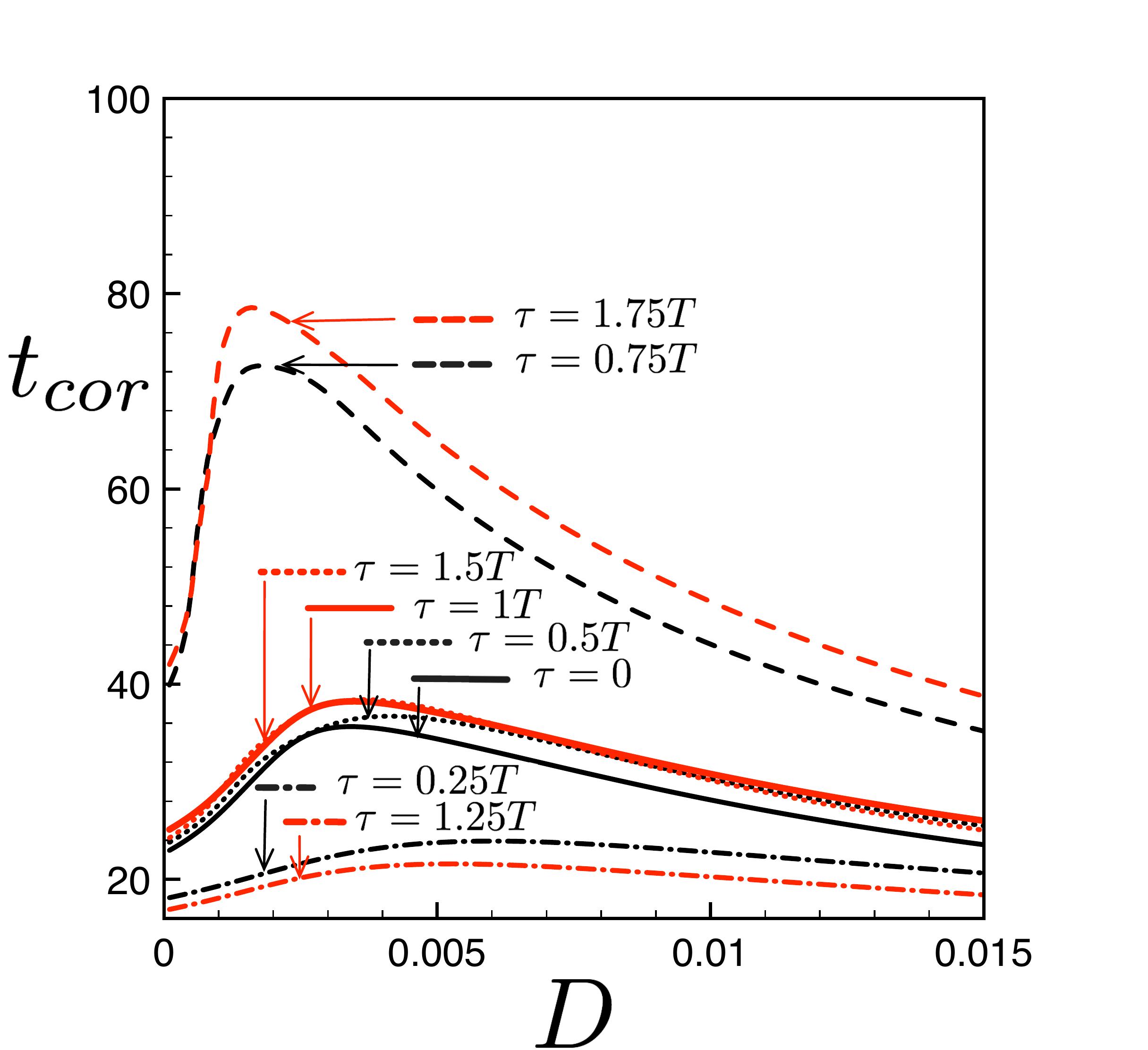}
\end{center}
\caption{Numerical results for the dependence of the correlation time on noise intensity $D$ for various values of time delay. Parameters: $K=0.024$, $\mu=0.5$, $\varepsilon = -0.06$, $\omega_0=1$, $T=2\pi$.}
\label{fig12}                                                                                                   
\end{figure}

The features of coherence resonance clearly show up in the plot of the correlation time versus the noise intensity. In analogy with Ref. \cite{GEF14}, we numerically calculate this dependence for different values of time delay (Fig. \ref{fig12}) to demonstrate the possibility of controlling coherence resonance by time-delayed feedback. Without time delay this dependence has a maximum at an intermediate noise intensity, which indicates coherence resonance. By properly choosing the delay time we can significantly enhance the effect of coherence resonance which becomes apparent for $\tau = (\frac{3}{4}+n)T$, $n\in \mathbb N$ (Fig. \ref{fig12}). For $\tau = (\frac{1}{4}+n)T$, $n\in \mathbb N$, no distinct  maximum is observed, thus coherence resonance is suppressed. In similarity to the probability distributions, the dependence of the correlation time upon noise intensity also clearly indicates that time delay has no impact on the system for $\tau = \frac{n}{2}T$, $n\in \mathbb N$. This can be understood from Fig.~\ref{fig3} and Eq.~(\ref{eq17}), which shows that the maximum shift of the saddle-node bifurcation points to larger and to smaller values of $\varepsilon$ occurs for
$\tau = (\frac{1}{4}+n)T$ and $\tau = (\frac{3}{4}+n)T$, respectively, 
while there is no shift for $\tau= \frac{n}{2}T$. The bimodality of the amplitude probability distribution, and hence coherence resonance, is enhanced if the operating point is brought closer to the saddle-node bifurcation of limit cycles.

%

\section{Conclusion}

We have analyzed the generalized Van der Pol oscillator in the presence of noise and delay. This model exhibits non-excitable dynamics, and in the regime of subcritical Hopf bifurcation it demonstrates the phenomenon of coherence resonance. The amplitude probability density distributions for various values of the time delay, and the correlation time in dependence upon time delay and noise intensity 
are studied to demonstrate the controllability of coherence resonance by delay. We find the suppression of noise-induced oscillations for $\tau=(\frac{1}{4}+n)T$, $n\in \mathbb N$ and their enhancement for $\tau=(\frac{3}{4}+n)T$,  $n\in \mathbb N$. The delay time values $\tau = \frac{1}{2}nT$,  $n\in \mathbb N$ do not have any impact on the coherence resonance. 

Coherence resonance in non-excitable systems in the presence of time delay has recently been considered theoretically in a model of Stuart-Landau oscillator \cite{GEF14}. The present study discloses the possibility to control coherence resonance by time delayed feedback in the experiment on an electronic circuit, which is especially relevant from the application point of view.

 \section{Acknowledgements}
This work was supported by DFG in the framework of SFB 910 and by the Russian Ministry of Education and
Science (project code 1008).

\bibliography{ref}
\bibliographystyle{prwithtitle}

\end{document}